\documentclass[prl,aps,twocolumn,showpacs]{revtex4}
\usepackage{epsfig}

\begin{document}

\author{Li-Bin Fu}
\affiliation{Institute of Applied Physics and Computational Mathematics, P.O. Box 8009
(28), 100088 Beijing, China, and \\
Max-Planck-Institute for the Physics of Complex systems N\"{o}thnitzer
Strasse 38, 01187 Dresden, Germany}
\title{Nonlocal effect of bipartite system induced by local cyclic operation}

\begin{abstract}
The state of a bipartite system may be changed by a cyclic operation applied
on one of its subsystem. The change is a nonlocal effect, and can be
detected only by measuring the two parts jointly. By employing the
Hilbert-Schmidt metric, we can quantify such nonlocal effects via measuring
the distance between initial and final state. We show that this nonlocal
property can be manifested not only by entangled states but also by the
disentangled states which are \textit{classically correlated}. Furthermore,
we study the effect for the system of two qubits in detail. It is
interesting that the nonlocal effect of disentangled states is limited by $1/%
\sqrt{2}$, while the entangled states can exceed this limit and reach $1$
for maximally entangled states.
\end{abstract}

\maketitle


Entanglement is a striking feature of composite quantum system, which has no
classical analog. Historically, since Einstein, Podolsky, and Rosen (EPR)
published their famous gedanken experiment in 1935 \cite{epr}, entanglement
had become a key issue in the debate about the foundations and
interpretation of quantum mechanics. The appeal was changed dramatically in
1964 by John Bell's theorem \cite{bell}. Bell inequalities \cite%
{bellnew,bellnew1,bellnew2} bound the correlations within any local and
realistic theory. According to Bell's theorem, there are some states of
composite system, when measurements are performed on the two subsystems
separated in space their results are correlated in a manner which cannot be
explained by local hidden variables models. For a quite long time,
entanglement was widely believed to be equivalent to the violation of a Bell
inequality. Whereas until 1989, Werner proved that even if Bell's inequality
is satisfied by a given composite system, there is no guarantee that its
state can be prepared by two distant observers who receive instructions from
a common source \cite{wer}. Thereafter, it is generally recognized in the
community that a quantum state of a system composed of two subsystems is
called entangled if and only if it is not a separable state, i.e. it can not
be expressed as
\begin{equation}
\sigma _{s}=\sum_{l}p_{l}\left| \psi _{A}^{l}\right\rangle \left\langle \psi
_{A}^{l}\right| \otimes \left| \psi _{B}^{l}\right\rangle \left\langle \psi
_{B}^{l}\right| .  \label{disa}
\end{equation}
where $p_{l}$ are positive real numbers and $\sum_{l}p_{l}=1.$\ A separable
system always satisfies Bell inequality, but the converse is only true for
pure states.

Nowadays, quantum entanglement has become not only a tool for exposing the
weirdness of quantum mechanics \cite{epr,bell}, but also a more powerful
resource in a number of applications \cite{myd1,myd2,myd3,myd4}. One of the
most important problem is the characterization and classification of mixed
entangled states. The most prominent criterion for deciding whether a given
state is entangled or not is known as positive partial transpose (PPT) test %
\cite{pptm}. For systems consisting with two qubits or a qubit and a qutrit,
PPT test is the necessary and sufficient conditions for presence of
entanglement. For systems of more than three parties, or for higher
dimensions system, the PPT test is only a sufficient criterion, since there
exist PPT-entangled states \cite{pptent}.

Entanglement witness (EW) are operators that are designed to detect presence
of entanglement in a state \cite{pptent,ew1,ew2}. A Hermitian operator $W$
is called entanglement witness if it has a positive expectation value for
all separable states, $Tr[W\sigma _{s}]\geq 0,$ while there exists at least
one state $\rho $ that $Tr[W\rho ]<0$ . Therefore, the state with negative
expectation should be entangled and it is said to be detected by the witness
$W.$ Entanglement witness is an important concept and provide a very useful
tool for experimental detection of entanglement \cite{cit23,cit24}.

In this letter, we generalize the nonlocal effect manifested by the
maximally entangled state in the quantum dense coding process \cite{myd3} to
any state of a bipartite system. By employing the Hilbert-Schmidt distance %
\cite{myd4}, we quantify this nonlocal effect. We find such nonlocal effect
vanishes for product states but not vanishes for \textit{classically
correlated }states (the disentangled states which can not be factorized) %
\cite{wer}. Furthermore, we investigate this effect for two qubits in
detail. The interesting thing is that the nonlocal effect of disentangled
states is bounded by $1/\sqrt{2}$, but for entangled states it can exceed
this limit and reach $1$ for maximally entangled states. Hence, the nonlocal
effect can be used to detect entanglement for some states.

At first, let us remind the reader of the dense coding process. Quantum
dense coding \cite{myd3} enables the communication of two bits of classical
information by transferring one qubit between two parties who share a
maximally entangled pair. At the beginning, one party, ''Alice'', \ prepares
a maximally entangled pair and sends one of the particles to another party,
''Bob''. Bob applies one of four possible unitary operations, and sends it
back to Alice. By measuring the two particles jointly, the outcomes of these
measurements tell her which of the four operations Bob applied and the
corresponding two-bit classical number.

In the quantum dense coding process, the subsystem of the treated particle
is not changed by the local unitary operation (or in other words, the
marginal statistics of measurements of the treated particle is unperturbed
after the local operation applied by Bob \cite{myd3}). The untreated
particle is fixed all the time. So, the states of both two subsystems are
not changed after the local unitary operation. However, the state of the
whole system is changed after the operation applied by Bob. The shift of the
state of the whole system is a nonlocal effect, since it can be observed
only by measuring the two particles jointly.

Now, let us consider more general cases. Assuming Alice and Bob share a
system compounded by two particles $A$ (in Alice's hand) and $B$ (in Bob's
hand), which is in a state described by the density operator $\rho $. The
subsystems are described by the reduced density operators, $\rho
_{0}^{A}=tr_{B}(\rho _{0})$ and $\rho _{0}^{B}=tr_{A}(\rho _{0})$
respectively. Bob applies a local unitary operation $U^{B}$ on the particle
in his hand which satisfies
\begin{equation}
\lbrack \rho _{0}^{B},U^{B}]=0.  \label{ub}
\end{equation}%
Obviously, the subsystem is not changed by such an operation. However, the
whole system will not always return to its initial state, i.e., $\rho
_{0}\neq \left( I\otimes U^{B}\right) \rho _{0}\left( U^{B\dagger }\otimes
I\right) $ in general. The change between the final and initial states can
not be detected locally. For convenience, we denote the operation satisfies
condition (\ref{ub}) as a local cyclic operation.

To denote the difference between the initial and final states, we introduce
the distance between two states \cite{myd4}. Here, we employ the
Hilbert-Schmidt metric, $D(\rho _{1}||\rho _{2})=Tr|\rho _{1}-\rho
_{2}|^{2}, $ to measure the distance between quantum states $\rho _{1}$ and $%
\rho _{2}$, where $|X|=\sqrt{X^{+}X}$. The Hilbert-Schmidt metric $D(\rho
_{1}||\rho _{2})\geq 0$ with the equality saturated iff $\rho _{1}=\rho _{2}$
\cite{vv98,cw99}. Then, we quantify the shift between the initial state $%
\rho _{0} $ and the final state $\rho _{f}=\left( I\otimes U^{B}\right) \rho
_{0}\left( U^{B\dagger }\otimes I\right) $ by
\begin{equation}
d(\rho _{0},U^{B})=\sqrt{D(\rho _{0}||\rho _{f})/2}.  \label{aaa}
\end{equation}%
By considering $Tr(\rho _{0}^{2})=Tr(\rho _{f}^{2})$, we can obtain
\begin{equation}
d(\rho _{0},U^{B})=\sqrt{Tr(\rho _{0}^{2})-Tr(\rho _{0}\rho _{f})}.
\label{dr}
\end{equation}%
Obviously, $d(\rho _{0},U^{B}(\tau ))\leq 1$ and $d(\rho _{0},U^{B})=1$ only
when the initial state is a pure state and it is orthonormal with the final
state. \ In fact, for $\rho _{0}=\left| \psi \right\rangle \left\langle \psi
\right| $ is a pure state, we can have $d(\rho _{0},U^{B})=\sqrt{1-F(\rho
_{0},\rho _{f})},$where $F(\rho _{0},\rho _{f})=\left\langle \psi \right|
\rho _{\tau }\left| \psi \right\rangle $ is just the Bures fidelity \cite%
{myd4}.

Therefore, we have $0\leq d(\rho _{0},U^{B}(\tau ))\leq 1$, and the equality
on the left is saturated iff $\rho _{0}=\rho _{f}.$ Hence, $d(\rho
_{0},U^{B}(\tau ))$ can be used to quantify the nonlocal shift of the state
induced by the local cyclic operation. For convenience, we use $d_{\max
}(\rho _{0})$ to denote the maximum value of $d(\rho _{0},U^{B})$ over all
the local operations $U^{B}$ which satisfy (\ref{ub}).

A state of a bipartite system can be written in the following form
\begin{widetext}
\begin{eqnarray}
\rho _{0} &=&\frac{1}{N_{A}N_{B}}\left[ I^{A}\otimes I^{B}+\sqrt{\frac{%
N_{A}(N_{A}-1)}{2}}\mathbf{r}^{A}\cdot \vec{\lambda}^{A}\otimes
I^{B}+\right.   \nonumber \\
&&\left. \sqrt{\frac{N_{B}(N_{B}-1)}{2}}I^{A}\otimes
\mathbf{r}^{B}\cdot
\vec{\lambda}^{B}+\sqrt{\frac{N_{A}(N_{A}-1)N_{B}(N_{B}-1)}{4}}\beta
_{ij}\lambda _{i}^{A}\otimes \lambda _{j}^{B}\right] , \label{rab}
\end{eqnarray}
\end{widetext}where $N_{A}$ and $N_{B}$ are the dimensions of each
subsystems, $\vec{\lambda}^{A}=(\lambda _{i}^{A};i=1,2,\cdots ,N_{A}^{2}-1)$
and $\vec{\lambda}^{B}=(\lambda _{i}^{A};i=1,2,\cdots ,N_{B}^{2}-1)$ are the
generators of $SU(N_{A})$ and $SU(N_{B})$ respectively, $\mathbf{r}%
^{A}=(r_{i}^{A};i=1,2,\cdots ,N_{A}^{2}-1)$ and $\mathbf{r}%
^{B}=(r_{i}^{B};i=1,2,\cdots ,N_{B}^{2}-1)$ are two Bloch vectors, and $%
\beta _{ij}$ are $(N_{A}^{2}-1)(N_{B}^{2}-1)$ real numbers which constructs
the so-called correlation matrix $\beta =\{\beta _{ij}\}$. The states of the
two subsystems are described by the following reduced density operators,
\begin{eqnarray}
\rho _{0}^{A} &=&\frac{1}{N_{A}}\left[ I^{A}+\sqrt{\frac{N_{A}(N_{A}-1)}{2}}%
\mathbf{r}^{A}\cdot \vec{\lambda}^{A}\right] ,  \nonumber \\
\rho _{0}^{B} &=&\frac{1}{N_{B}}\left[ I^{B}+\sqrt{\frac{N_{B}(N_{B}-1)}{2}}%
\mathbf{r}^{B}\cdot \vec{\lambda}^{B}\right] .  \label{sub}
\end{eqnarray}
It is easy to see $\rho _{f}^{B}=$ $U^{B}\rho _{0}^{B}U^{B+}=\rho _{0}^{B}$
for the local cyclic operation defined by (\ref{ub})$.$ An interesting case
is for the states of which $|\mathbf{r}^{A}|=|\mathbf{r}^{B}|=0.$ For such
states, any local unitary operation is a local cyclic operation. The
maximally entangled states and the Werner states \cite{wer} are belong to
this case.

With the condition (\ref{ub}), and the trace relation $Tr(\lambda
_{i}\lambda _{j})=2\delta _{ij}$ for the generators of $SU(N),$ one can
obtain
\begin{equation}
d(\rho _{0},U^{B})=\sqrt{\frac{(N_{A}-1)(N_{B}-1)}{N_{A}N_{B}}\left( |\beta
|^{2}-\sum_{i,j}\beta _{ij}\beta _{ij}^{f}\right) },  \label{dd}
\end{equation}
in which $|\beta |^{2}=\sum_{i,j}\beta _{ij}\beta _{ij}$ and $\beta
_{ij}^{f} $ are the elements of correlation matrix of the final state, which
are defined by the following relations
\begin{equation}
\beta _{ij}^{f}\lambda _{i}^{A}\otimes \lambda _{j}^{B}=\beta _{ij}\lambda
_{i}^{A}\otimes U^{B}\lambda _{j}^{B}U^{B^{+}}.  \label{reb}
\end{equation}
In the above calculation, we have used the relation $|\beta |=|\beta ^{f}|$.
If regard the expression $\sum_{i,j}\beta _{ij}\beta _{ij}^{f}$ as the inner
product of two vectors, so $\sum_{i,j}\beta _{ij}\beta _{ij}^{f}\leq |\beta
|^{2}.$ Then we can easily prove that $d(\rho _{0},U^{B})\geq 0$ and $d(\rho
_{0},U^{B})=0$ if an only if $\rho _{f}=\rho _{0}.$

\textit{Theorem.} For the state (\ref{rab}) of which $\beta _{ij}=\alpha
r_{i}^{A}r_{j}^{B}$ $(0\leq \alpha \leq 1),$ $d_{\max }(\rho _{0})=0$, i.e.,
such state can not have the nonlocal shift induced by a local cyclic
operation.

\textit{Proof.} From Eq. (\ref{sub}), the condition , $[\rho
_{0}^{B},U^{B}]=0$, is equivalent to
\begin{equation}
\mathbf{r}^{B}\cdot \vec{\lambda}^{B}=U^{B}\left( \mathbf{r}^{B}\cdot \vec{%
\lambda}^{B}\right) U^{B^{+}}.  \label{con}
\end{equation}
If $\beta _{ij}=\alpha r_{i}^{A}r_{j}^{B}$ $(0<\alpha \leq 1),$ the
correlation matrix $\beta =\alpha \mathbf{r}^{A}\cdot \vec{\lambda}%
^{A}\otimes \mathbf{r}^{B}\cdot \vec{\lambda}^{B}.$ Then, $\beta ^{f}=\alpha
\mathbf{r}^{A}\cdot \vec{\lambda}^{A}\otimes U^{B}\mathbf{r}^{B}\cdot \vec{%
\lambda}^{B}U^{B^{+}}.$ From (\ref{con}), we obtain $\beta ^{f}=\beta ,$
i.e., $\beta _{ij}^{f}=\alpha r_{i}^{A}r_{j}^{B}=\beta _{ij}.$ Hence, $%
d(\rho _{0},U^{B})=0$. The proof is end.

From this theorem, we know that the nonlocal shift can not be observed for
any product state $(\rho =\rho ^{A}\otimes \rho ^{B}).$ Hence, the effect
can not be observed for the disentangled pure states since they are product
states.

It is well-known that some disentangled mixed states are able to exhibit
non-locality \cite{wer}, which are so-called \textit{classically correlated}
states . A property of this nonlocal effect is that the effect can be
observed for the disentangled states which are \textit{classically correlated%
}. A state $\rho $ is classically correlated if it can be expressed as
\begin{equation}
\rho =\sum_{l}^{M}p_{l}\left| \psi _{A}^{l}\right\rangle \left\langle \psi
_{A}^{l}\right| \otimes \left| \psi _{B}^{l}\right\rangle \left\langle \psi
_{B}^{l}\right| .  \label{dis}
\end{equation}
with $M>1,$ where $p_{l}$ are positive real numbers and $%
\sum_{l}^{M}p_{l}=1. $ Denoting $\mathbf{r}^{A_{l},B_{l}}$ as the Bloch
vectors corresponding to $\left| \psi _{A}^{l}\right\rangle \left\langle
\psi _{A}^{l}\right| $ and $\left| \psi _{B}^{l}\right\rangle \left\langle
\psi _{B}^{l}\right| $ respectively. The Bloch vectors for such state are $%
\mathbf{r}^{A}=\sum_{l}^{M}p_{l}\mathbf{r}^{A_{l}}$ and $\mathbf{r}%
^{B}=\sum_{l}^{M}p_{l}\mathbf{r}^{B_{l}},$ and $\beta
_{ij}=\sum_{l}^{M}p_{l}r_{i}^{A_{l}}r_{j}^{B_{l}}.$ Then form Eq. (\ref{reb}%
) we can know that, for a local operation $U^{B}$ satisfying (\ref{con}), $%
\beta _{ij}^{f}\neq \beta _{ij}$ unless $M=1$ (or $\mathbf{r}^{A_{1}}=%
\mathbf{r}^{A_{2}}=\cdots =\mathbf{r}^{A_{M}}$)$.$Therefore, this nonlocal
effect can be observed for the disentangled states which are \textit{%
classically correlated}.

To make the above discussion more clear, we study this nonlocal effect for
two qubits in detail. For qubits, it is common to choose the generators of $%
SU(2)$ as Pauli matrices, i.e., $\sigma _{1}=\left(
\begin{array}{cc}
0 & 1 \\
1 & 0%
\end{array}
\right) ,$ $\sigma _{2}=\left(
\begin{array}{cc}
0 & -i \\
i & 0%
\end{array}
\right) ,$ and $\sigma _{3}=\left(
\begin{array}{cc}
1 & 0 \\
0 & -1%
\end{array}
\right) .$ The unitary operation applied on the subsystem can be expressed
as $U^{B}=e^{i\frac{\varphi }{2}\mathbf{u}\cdot \vec{\sigma}}$ where $%
\mathbf{u}$ is a unit vector. At first, we discuss the case for pure states.
For notational convenience, we assume the initial state as follows
\begin{equation}
|\psi \rangle =k_{1}\left| 00\right\rangle +k_{2}\left| 11\right\rangle ,
\label{state}
\end{equation}
with $|k_{1}|^{2}+|k_{2}|^{2}=1.$ For $|k_{1}|=|k_{2}|=\frac{\sqrt{2}}{2},$
the state is a maximally entangled state. The states of two subsystems are
\begin{equation}
\rho _{A}=\rho _{B}=\frac{1}{2}(\left[ I+\left(
|k_{1}|^{2}-|k_{2}|^{2}\right) \sigma _{3}\right] ).  \label{ras}
\end{equation}
It is easy to prove that the unitary operation which satisfies (\ref{ub})
can be expressed as
\begin{equation}
U^{B}=e^{i\varphi /2\sigma _{3}}.  \label{us}
\end{equation}
Then, $\rho _{f}=|\psi _{f}\rangle \left\langle \psi _{f}\right| $ with $%
|\psi _{f}\rangle =I\otimes U^{B}|\psi \rangle $. From Eq. (\ref{dr}), we
obtain
\begin{equation}
d(\psi ,U^{B})=2|k_{1}k_{2}\sin \varphi /2|.  \label{ddd}
\end{equation}
Obviously, $d_{\max }(\psi )=2|k_{1}k_{2}|,$ which just equals to the degree
of entanglement for pure state of two qubits suggested in Refs. \cite%
{fude,deg,woot}. The definition of entanglement degree consists with the
violation of Bell inequality. The optimal form of Bell inequality for the
entangled qubits is known as the Clauser-Horne-Shimony-Holt (CHSH)
inequality \cite{bellnew}. It has been shown by Gisin \cite{gisin} that any
entangled pure state of qubit pair can violate the CHSH inequality and the
maximum violation is $B_{\max }(\psi )=2\sqrt{1+4|k_{1}k_{2}|^{2}}.$
Obviously, $B_{\max }(\psi )=2\sqrt{1+d_{\max }^{2}(\psi )}.$ Therefore, the
nonlocal effect can be used to quantify the entanglement of pure state of
qubit pair.

Although disentangled states may have such nonlocal effect, the
maximum value for disentangled states is bounded and this boundary
can be exceeded by
entangled states. For the disentangled states expressed by Eq. (\ref{dis}), $%
\beta _{ij}=\sum_{l}^{M}p_{l}r_{i}^{A_{l}}r_{j}^{B_{l}}.$ Then, one can have
$|\beta |^{2}\leq 1$ since $|\mathbf{r}^{A_{l}}|=|\mathbf{r}^{B_{l}}|=1.$ On
the other hand, $\sum_{i,j}\beta _{ij}\beta _{ij}^{f}\geq -|\beta |^{2}$ for
qubit \cite{fucite}. Therefore, from Eq. (\ref{dd}) we can immediately
obtain
\begin{equation}
d_{\max }\leq \frac{1}{\sqrt{2}}  \label{class}
\end{equation}%
for the states which are \textit{classically correlated}. Therefore, the
nonlocal shift of disentangled states is bounded by $1/\sqrt{2}.$

From (\ref{ddd}) we know that the shifts of entangled states can exceed this
limit and reach $1$ for maximally entangled states. It is interesting that
the entangled states violate the \textit{classically correlated} states by
the factor $\sqrt{2},$ which consists with the CHSH inequality.

Any state violating the inequality (\ref{class}) is entangled.
Therefore, the nonlocal effect can be employed to detect
entanglement of some states. On the other hand, because $d_{\max
}=0$ for the product states, so we can use this nonlocal effect to
identify product states.

It is not difficult to observe this nonlocal effect by using the
following Bell type experiment.

Under the transformation $U^{B}=e^{i\frac{\varphi
}{2}\mathbf{u}\cdot \vec{\sigma}}$ , we can get
\begin{eqnarray}
\sigma _{i}^{f} &=&U^{B}\sigma _{i}U^{B\dagger }=\cos \varphi \sigma _{i}
\nonumber \\
&&+\epsilon _{ijk}u_{j}\sin \varphi \sigma _{k}+2\sin ^{2}\frac{\varphi }{2}%
u_{i}\mathbf{u}\cdot \vec{\sigma}.  \label{sg}
\end{eqnarray}%
In fact, $\mathbf{\sigma }^{f}=(\sigma _{1}^{f},\sigma _{2}^{f},\sigma
_{3}^{f})$ is just another set of pauli matrices. From (\ref{reb}), we have
\begin{equation}
\beta _{ij}^{f}\sigma _{i}\otimes \sigma _{j}=\beta _{ij}\sigma _{i}\otimes
\sigma _{j}^{f}.  \label{bbt}
\end{equation}

Let us perform the measurements either $A_{1}$ or $A_{2}$ on one particle,
and either $B_{1}$ or $B_{2}$ on the other, where $A_{1}=\mathbf{n}^{1}\cdot
\mathbf{\vec{\sigma}}$, $A_{2}=\mathbf{n}^{2}\cdot \mathbf{\vec{\sigma}}$, $%
B_{1}=\mathbf{m}^{1}\cdot \mathbf{\vec{\sigma}}$, and $B_{2}=\mathbf{m}%
^{2}\cdot \mathbf{\vec{\sigma}}$. Let $E(A,B),$ denote the quantum
expectation value of the product $AB$. We define $F$ as
\begin{eqnarray}
F &=&E(A_{1},B_{1})+E(A_{1},B_{2})  \nonumber \\
&&+E(A_{2},B_{1})-E(A_{2},B_{2}),  \label{ssdd}
\end{eqnarray}
which is just the CHSH expression \cite{bellnew}. Let us introduce the
measurement matrix $T$ as $%
T_{ij}=(n_{i}^{1}+n_{i}^{2})m_{j}^{1}+(n_{i}^{1}-n_{i}^{2})m_{j}^{2},\;%
\;i,j=1,2,3.$ We can obtain the quantum expectation of $F$ for the initial
state $\rho _{0}$,
\begin{equation}
F(\rho _{0},T)=\sum\limits_{i,j}\beta _{ij}T_{ij}.  \label{fa}
\end{equation}

Then, for the final state $\rho _{f},$ if one chooses the measurements, $%
A_{1}^{f}=A_{1}$, $A_{2}^{f}=A_{2}$, $B_{1}^{f}=\mathbf{m}^{1}\cdot \vec{%
\sigma}^{f}$, and $B_{2}^{f}=\mathbf{m}^{2}\cdot \vec{\sigma}^{f},$ we can
prove that
\begin{equation}
F(\rho _{0},T)=F(\rho _{f},T^{f}),  \label{eq}
\end{equation}
in which $T^{f}$ is the measurement matrix corresponding to the measurement
settings for the final state.

At first, we let $B_{1}=\sigma _{1}$, and $B_{2}=\mathbf{\sigma }_{2}$ be
fixed, and then change the settings for $A_{1}$ and $A_{2}$ to find the
maximal value $F_{\max }(\rho _{0},T)$ for the initial state. We can obtain
the optimal settings $\overline{A}_{1}$ and $\overline{A}_{2}.$

Secondly, we apply measurements on the final state. At this time, we let $%
A_{1}^{f}=\overline{A}_{1}$ and $A_{2}^{f}=\overline{A}_{2}$ be fixed. Then
we vary the settings for the others. From the above discussion, we can know
that $F(\rho _{f},T^{f})$ will reach its maximal value (which must equal to $%
F_{\max }(\rho _{0},T)$)$,$ if $B_{1}^{f}=\sigma _{1}^{f}$, and $%
B_{2}^{f}=\sigma _{2}^{f}.$ So, from the relations between pauli matrices,
one can get $\sigma _{3}^{f}$. Hence, from (\ref{dd}) and (\ref{bbt}) we can
obtain $\beta ^{f}$ and the $d(\rho _{0},U^{B})$ immediately.

In conclusion, we have investigated nonlocal effects for the bipartite
system induced by local cyclic operations of one of its subsystem. We employ
the Hilbert-Schmidt distance to measure the nonlocal effect. Such nonlocal
shifts vanish for product states, but do not vanish for disentangled states
that are only \textit{classically correlated}. Therefore this nonlocal
effect can be used classify the disentangled states. For qubit pair, we show
that the nonlocal shift of disentangled states is limited by $1/\sqrt{2}$,
while the shifts of entangled states can exceed this limit and reach $1$ for
maximally entangled states. Hence, the nonlocal effect can be used as a
sufficient condition of detecting entanglement.

In fact, the nonlocality is due to the existence of correlations in compound
quantum systems, which is more general notion than entanglement. It is
well-known that the local operations on the subsystem of the compound
quantum system in the distance labs paradigm can produce nonlocal
consequences. In this letter attention was focused on nonlocal properties
caused by the local operations which do not make the subsystem changed. Such
nonlocal property is not equivalent to entanglement in general. We hope that
such nonlocal property, especially, the fact that nonlocal property implied
by disentangled states, will draw much more attention of physicists on
studying nonlocality and entanglement of quantum systems.

This work was supported by National Nature Science Foundation of China
(10445005,10474008), the Science and Technology fund of CAEP, and the
National fundamental Research Programme of China (Grant No. 2005CB3724503).
Dr. L.-B Fu acknowledges the support of the Alexander von Humboldt
Foundation.


\begin{thebibliography}{99}
\bibitem{fucite} This expression is not ture for higher dimensional systems,
because for the higher dimensional systems, the Hilbert space is asymmetry
for the Bloch vectors \cite{fu4d}.
\end{thebibliography}
\end{document}